\documentclass[11pt]{article}
\usepackage[margin=1in]{geometry}
\usepackage{amsmath,amssymb,amsthm,mathtools,booktabs}
\usepackage[hidelinks]{hyperref}
\usepackage{enumitem}
\usepackage{microtype}
\newtheorem{theorem}{Theorem}
\newtheorem{lemma}[theorem]{Lemma}
\newtheorem{corollary}[theorem]{Corollary}

\newtheorem{observation}[theorem]{Observation}
\newcommand{\F}{\mathbb F}
\newcommand{\Z}{\mathbb Z}
\newcommand{\Q}{\mathbb Q}
\newcommand{\SAT}{\ensuremath{\#\mathrm{3SAT}}}
\newcommand{\poly}{\operatorname{poly}}
\title{Subexponential-Time Provers for Sparse \SAT\thanks{This is a preliminary draft. A revised and extended version will follow.}}
\author{Vaani Goenka\\Vachani School of Advanced Computing\\Department of Computer Science, Ashoka University \and Mahabir Prasad Jhanwar\\Vachani School of Advanced Computing\\Department of Computer Science, Ashoka University}
\date{November 28, 2026}
\begin{document}
\maketitle
\begin{abstract}
The classical sum-check protocol for \(\#\mathrm{SAT}\) has a polynomial-time verifier, but the standard honest prover runs in exponential time. The recent subexponential-time algorithms for \(\#\mathrm{SAT}\) on bounded-width CNF formulae raise a natural question: can an honest prover exploit these algorithms without sacrificing efficient, information-theoretic verification? A faster Boolean counting algorithm is not known to give a faster prover by itself, because the verifier's challenges are field elements, and the prover's messages are weighted sums over Boolean assignments rather than counts. We observe that these weighted sums retain a bounded-arity product structure and can be evaluated using the generalized counting machinery in the recent ETH refutations. As a direct application, we obtain an honest prover that runs in time \(2^{O(n/\sqrt{\log n})}\poly(n)\) for sparse 3-CNF formulae. Moreover, since the sum-check protocol itself remains unchanged, the verifier's running time, communication complexity, and information-theoretic soundness stay the same. 
\end{abstract}

\section{Introduction}
Interactive proofs separate the cost of computing an answer from the cost of verifying it. The sum-check protocol of Lund, Fortnow, Karloff, and Nisan~\cite{lfkn} is a classical interactive proof protocol.

Given a Boolean formula \(\phi\), we want to verify the claim that \(\phi\) has exactly \(K\) satisfying assignments. Unlike satisfiability, where a satisfying assignment is a polynomial-size certificate, no polynomial-size certificate for such a claim is known; one would imply
\(\mathsf{NP} = \mathsf{coNP}\). In the sum-check protocol, \(\phi\) is
arithmetized into a polynomial over a finite field. The number of Boolean points at which this polynomial evaluates to \(1\) equals the number of satisfying assignments of \(\phi\). The protocol yields a polynomial-time verifier by reducing the verifier's work to evaluating this polynomial at a single point of the field. Thaler~\cite[Section~4.2]{thaler} bounds the running time of the standard honest prover by \(O(S^2 \cdot 2^n)\), where \(S\) is the size of the formula and \(n\) is its number of variables.

The recent algorithmic results of OpenAI~\cite{openaieth} and Anthropic~\cite{claudeeth} refute the counting exponential-time hypothesis (\(\#\mathrm{ETH}\)). In particular, on sparse 3-CNF instances (\(O(n)\) clauses), they state that the complexity of \SAT\ is bounded by \(2^{O(n/\sqrt{\log n})}\poly(n)\). A natural question is whether the cost of the prover in classical sum-check can also be reduced to the same asymptotic scale without changing the protocol itself, so that its polynomial-time, information-theoretic verification is kept.

This is not resolved by directly applying the faster counting algorithm. The verifier's challenges are field elements, so the honest
prover's messages are weighted sums over Boolean assignments rather than counts of satisfying assignments. The observation we use is that the 3-CNF arithmetization has the bounded-arity product structure needed by the generalized counting routines in the ETH-refutation preprints.

\paragraph{Contributions and scope.} 
We define the Weighted Counting problem
(Section~\ref{sec:weighted}) and reduce each of the honest prover's messages to instances of it. Only the prover's computation changes, and the sum-check protocol itself remains unchanged. We then show how the counting routines in the recent ETH algorithms solve the Weighted Counting problem. For sparse 3-CNF formulae, this gives an honest prover running in time \(2^{O(n/\sqrt{\log n})}\poly(n)\). The reduction applies to 3-CNF formulae of any density, but a subexponential prover is obtained only for sparse 3-CNF formulae. 

\section{Preliminaries}\label{sec:prelim}
Write \([n]=\{1,\ldots,n\}\). The \(\#\mathrm{SAT}\) problem is the following:

\paragraph{Input.} A CNF formula \(\phi=\bigwedge_{j=1}^{m} C_j\) over
\(x_1,\ldots,x_n\), where each clause \(C_j\) is a disjunction of literals, each
a variable \(x_\ell\) or its negation \(\neg x_\ell\), with no variable
occurring twice in a clause.

\paragraph{Output.} The number \(\#\phi\) of assignments \(x\in\{0,1\}^n\) that
satisfy \(\phi\).
\bigskip
\newline 
The \SAT\ problem is the restriction of \(\#\mathrm{SAT}\) to 3-CNF formulae,
in which each clause has at most three literals. Throughout, \(n\) is the
number of declared Boolean variables, \(m\) the number of clauses, and \(p\) a
prime of polynomial bit length. We work in \(\F_p\), where
\(p>\max\{2^n,3m/\varepsilon\}\) for target soundness \(\varepsilon\).


\paragraph{The classical sum-check protocol:} 
In the sum-check protocol~\cite{lfkn}, a verifier checks a prover's claimed sum of a multivariate polynomial over all boolean assignments to its variables. Let \(g\) be a polynomial over
\(\F_p\) in the formal variables \(X=(X_1,\ldots,X_n)\), and let the prover's
claimed sum be \(K\). Then the verifier wishes to check
\begin{equation}\label{eq:claim}
 K=\sum_{x\in\{0,1\}^n}g(x) \quad\text{in }\F_p.
\end{equation}
At round \(i\), given the previous challenges \(r_1,\ldots,r_{i-1}\), the honest prover sends the `round polynomial'
\begin{equation}\label{eq:message}
 g_i(T)=\sum_{x\in\{0,1\}^{n-i}}
 g(r_1,\ldots,r_{i-1},T,x).
\end{equation}
This polynomial has degree at most \(\deg_{X_i} g\). The verifier checks consistency at each round by checking \(g_1(0)+g_1(1)=K\) in the first round and \(g_i(0)+g_i(1)=g_{i-1}(r_{i-1})\) in each subsequent round. In the final round it checks \(g_n(r_n)=g(r_1,\ldots,r_n)\). 

\paragraph{Arithmetization.} Since \(\#\mathrm{SAT}\) requires us to count
satisfying assignments, that is, to compute \(\sum_{x\in\{0,1\}^n}\phi(x)\), it
lends itself well to the shape of sum-check~\eqref{eq:claim}. It suffices to
find a polynomial \(g\) over \(\F_p\) with \(g(x)=\phi(x)\) for every
\(x\in\{0,1\}^n\). We construct such a polynomial for 3-CNF formulae.

Let \(\phi=\bigwedge_{j=1}^{m} C_j\) be a 3-CNF formula over \(x_1,\ldots,x_n\).
The \emph{scope} of \(C_j\) is the set \(I_j\subseteq[n]\) of
indices \(\ell\) such that \(x_\ell\) or \(\neg x_\ell\) occurs in \(C_j\), so
\(|I_j|\le 3\). For \(x\in\{0,1\}^n\), write \(x|_{I_j}\) for the restriction of
\(x\) to \(I_j\). Whether \(x\) satisfies \(C_j\) depends only on \(x|_{I_j}\).

\bigskip
Over \(\F_p\), define for \(j\in[m]\) and \(\ell\in I_j\)
\[
  L_{j,\ell}(X_\ell)=
  \begin{cases}
    X_\ell,   & \text{if } x_\ell \text{ occurs in } C_j,\\
    1-X_\ell, & \text{if } \neg x_\ell \text{ occurs in } C_j.
  \end{cases}.
\]
The clause \(C_j\) is false exactly when all its literals are false. Accordingly,
the extension of \(C_j\) is the polynomial in \(X_{I_j}=(X_\ell)_{\ell\in I_j}\)
defined by
\begin{equation}\label{eq:clause}
  f_j(X_{I_j})=1-\prod_{\ell\in I_j}\bigl(1-L_{j,\ell}(X_\ell)\bigr).
\end{equation}
For \(u\in\{0,1\}^{I_j}\), \(f_j(u)=1\) if \(u\) satisfies \(C_j\) and
\(f_j(u)=0\) otherwise.
\bigskip
\newline
Now, we build the extension of \(\phi\) using the extensions of its clauses as follows:
\begin{equation}\label{eq:arith}
 g(X_1,\ldots,X_n)=\prod_{j=1}^m f_j(X_{I_j}).
\end{equation}
Then \(g(x)=\phi(x)\) for \(x\in\{0,1\}^n\), so the number of satisfying
assignments of \(\phi\) is
\begin{equation}\label{eq:count}
 \sum_{x\in\{0,1\}^n}\prod_{j=1}^{m} f_j(x|_{I_j}).
\end{equation}

Let \(d_i\) be the number of clauses whose scope contains \(i\), that is, the
number of occurrences of \(x_i\) or \(\neg x_i\) in \(\phi\).

\begin{observation}\label{obs:basic}
\begin{enumerate}[itemsep=0.5pt, label=(\roman*), ref=\theobservation(\roman*)]
  \item\label{obs:eval} The polynomial \(g\) can be evaluated at any point of
        \(\F_p^n\) with \(O(m)\) field operations.
  \item\label{obs:deg} The polynomial \(g\) has degree \(d_i\) in \(X_i\), so each
        round polynomial \(g_i\)~\eqref{eq:message} has degree at most \(d_i\).
  \item\label{obs:sum} The \(d_i\) are computable from \(\phi\) in \(O(m)\) time,
        and \(\sum_i d_i \le 3m\).
\end{enumerate}
\end{observation}

\paragraph{\SAT\ using sum-check.} An input to the sum-check protocol for
\SAT\ is a pair \((\phi,K)\), where \(K\in\{0,\ldots,2^n\}\) is the claimed
number of satisfying assignments of \(\phi\). To check \(K\), the verifier runs
the sum-check protocol on the arithmetization \(g\) of \(\phi\)
from~\eqref{eq:arith}. Since \(g(x)=\phi(x)\) for every \(x\in\{0,1\}^n\), the
sum in~\eqref{eq:claim} equals \(\#\phi\) modulo \(p\). Because \(K\) and
\(\#\phi\) both lie in \(\{0,\ldots,2^n\}\) and \(p>2^n\), the
claim~\eqref{eq:claim} holds exactly when \(K=\#\phi\).

The standard soundness analysis bounds the probability of accepting a false initial claim by \(\sum_i d_i/p\le 3m/p\). Completeness is perfect. The communication complexity is \(O(n+m)\) field elements. In the final round, the verifier computes \(g(r_1,\ldots,r_n)\) itself from \(\phi\) with \(O(m)\) field operations (Observation~\ref{obs:eval}), which brings its total cost to \(O(n+m)\) field operations.

The standard honest prover computes each round polynomial~\eqref{eq:message}
as a sum over all \(2^{n-i}\) Boolean assignments to the remaining variables.
Thaler~\cite[Section~4.2]{thaler} bounds its cost by \(O(S^2\cdot 2^n)\) field
operations, or \(2^{O(n)}\poly(n,m,\log p)\) bit operations, where \(S\) is the
size of \(\phi\). Up to polynomial factors, this is the time to solve
\(\#\mathrm{SAT}\) naively, by enumerating all \(2^n\) assignments. So the
natural question is whether a faster algorithm for \(\#\mathrm{SAT}\) gives a
faster honest prover.

\section{Bounded-Arity Integer Weighted Counting}\label{sec:weighted}
The Bounded-arity Integer Weighted Counting problem (Weighted Counting, for short) is defined as follows:
Fix a positive integer \(k\), the maximum number of variables in a constraint's scope.

\paragraph{Input.}
\begin{itemize}
\item A nonnegative integer \(N\) specifying the Boolean variables \(x_1,\ldots,x_N\).
\item A nonnegative integer \(M\) specifying the number of weighted constraints. For each \(j\in[M]\), the \(j\)-th constraint consists of a scope \(I_j\subseteq[N]\) of size at most \(k\), listed in increasing order, and an explicit table \(w_j:\{0,1\}^{I_j}\to\Z\). The table gives a signed binary integer for each of the \(2^{|I_j|}\) Boolean settings, in lexicographic order. Different constraints may have overlapping or
identical scopes. A constraint with an empty scope has one table entry.
\end{itemize}

\paragraph{Output.} Return the signed integer
\begin{equation}\label{eq:partition}
 W=\sum_{x\in\{0,1\}^N}\prod_{j=1}^{M} w_j(x|_{I_j}).
\end{equation}
Here \(x|_{I_j}\) is the restriction of \(x\) to \(I_j\), as in Section~\ref{sec:prelim}. The count~\eqref{eq:count} for a 3-CNF formula is the special case with \(k=3\), \(N=n\), \(M=m\), the clause scopes \(I_j\), and the tables \(w_j=f_j\) on \(\{0,1\}^{I_j}\). A constraint with an empty scope contributes a constant factor, and each
variable that occurs in no scope contributes a factor of~\(2\).

Let \(s=N+M\), and let \(B\) bound the binary length of
each weight. Each table has at most \(2^k\)
entries, but writing and processing those entries costs more as
\(B\) grows. The product for a single assignment has magnitude at
most \(2^{MB}\), and there are \(2^N\) assignments, so the exact
output has at most \(N+MB+O(1)\) bits. In the sum-check reduction
below, the least nonnegative representatives lie in \(\{0,\ldots,p-1\}\),
so \(B_p=\lceil\log_2 p\rceil+1\) is a valid bound even when the
number of remaining Boolean variables becomes small.

\begin{lemma}[Weighted Counting for one sum-check evaluation]\label{lem:reduction}
Given a 3-CNF formula \(\phi\), a round \(i\in[n]\), the earlier challenges \(r_1,\ldots,r_{i-1}\in\F_p\), and a point \(a\in\F_p\), one can construct a Weighted Counting instance with \(k=3\), \(N=n-i\), and \(M=m\). Its table entries have at most \(B_p\) bits. If the Weighted Counting algorithm returns \(W\), then \(g_i(a)=W\bmod p\). Constructing the input tables and reducing the output take \(\poly(n,m,\log p)\) bit operations.
\end{lemma}
\begin{proof}
Substitute \(r_1,\ldots,r_{i-1},a\) for the first \(i\) variables in each \(f_j\), and rename \(x_{i+1},\ldots,x_n\) as \(x_1,\ldots,x_{n-i}\). The \(j\)-th constraint gets the scope \(I_j'=\{\ell-i:\ell\in I_j,\ \ell>i\}\subseteq[n-i]\). For each of the at most eight Boolean settings \(u\) of \(I_j'\), evaluate the substituted extension in \(\F_p\) and store its least nonnegative integer representative as \(w_j(u)\). For a fixed assignment \(x\) to the remaining variables, multiplying the table entries gives the product of the extensions \(f_j\) at \((r_1,\ldots,r_{i-1},a,x)\). That product is \(g(r_1,\ldots,r_{i-1},a,x)\). Summing over \(x\) gives \(W\bmod p=g_i(a)\). Each clause table takes at most eight field evaluations to construct, and the integer sum has polynomial bit length by the bound above.
\end{proof}
\begin{lemma}[Weighted Counting for all round polynomials]\label{lem:rounds}
Given an algorithm for Weighted Counting, the honest
prover can compute all round polynomials \(g_1,\ldots,g_n\) using
\(\sum_{i=1}^{n}(d_i+1)\le n+3m\) calls to it, plus \(\poly(n,m,\log p)\)
additional bit operations.
\end{lemma}
\begin{proof}
The prover computes all \(d_i\) from \(\phi\) in \(O(m)\) time
(Observation~\ref{obs:sum}). Fix a round \(i\) and the challenges \(r_1,\ldots,r_{i-1}\). By Observation~\ref{obs:deg}, \(\deg g_i\le d_i\), so \(g_i\) is determined by its values at any \(d_i+1\) distinct points of \(\F_p\); such points exist since \(d_i\le 3m<p\). By Lemma~\ref{lem:reduction}, each value \(g_i(a)\) equals \(W \bmod p\) for one instance with \(k=3\), \(N=n-i\), \(M=m\), and entries of at most \(B_p\) bits, constructed in
\(\poly(n,m,\log p)\) bit operations. The prover makes one call per point and recovers the coefficients of \(g_i\) by interpolation over \(\F_p\) in
\(\poly(m,\log p)\) bit operations. Round \(i\) thus uses \(d_i+1\) calls, and \(\sum_i(d_i+1)=n+\sum_i d_i\le n+3m\) by Observation~\ref{obs:sum}.
\end{proof}

\paragraph{Example with five variables.}
Work over \(\F_{37}\) and take the three clauses
\[
 C_1=(x_1\lor x_2),\qquad
 C_2=(\neg x_1\lor x_3\lor x_4),\qquad
 C_3=(x_2\lor\neg x_3\lor x_4).
\]
The declared variable \(x_5\) is unused. In round \(2\), let \(r_1=2\).
To compute \(g_2(3)\), set \(X_2=3\). The extension of \(C_1\) gives the
constant \(-1\), stored as \(36\). The extension of \(C_2\) gives \(-1\) at
\((x_3,x_4)=(0,0)\) and \(1\) elsewhere. The extension of \(C_3\) gives \(3\) at
\((x_3,x_4)=(1,0)\) and \(1\) elsewhere. Thus \(N=M=3\), with empty
scope for \(C_1\) and scope \(\{3,4\}\) for both other constraints,
using the original indices.

\begin{center}
\begin{tabular}{@{}ccc rrr r@{}}
\toprule
\(x_3\) & \(x_4\) & \(x_5\) & \(w_1\) & \(w_2\) & \(w_3\) & Integer product\\
\midrule
0 & 0 & 0 & 36 & 36 & 1 & 1296\\
0 & 0 & 1 & 36 & 36 & 1 & 1296\\
0 & 1 & 0 & 36 & 1 & 1 & 36\\
0 & 1 & 1 & 36 & 1 & 1 & 36\\
1 & 0 & 0 & 36 & 1 & 3 & 108\\
1 & 0 & 1 & 36 & 1 & 3 & 108\\
1 & 1 & 0 & 36 & 1 & 1 & 36\\
1 & 1 & 1 & 36 & 1 & 1 & 36\\
\bottomrule
\end{tabular}
\end{center}

The constant weight appears in every row, and the two settings of
\(x_5\) give identical products. The weighted count is
\(W=2(1296+36+108+36)=2952\), so \(g_2(3)=29\) in \(\F_{37}\).

Since \(x_2\) occurs in two clauses, \(g_2(T)\) has degree at most two.
The prover can use the three evaluation points \(0,1,3\). At \(a=0\)
and \(a=1\), the corresponding tables give the integer sums
\(2(2\cdot 36+2+0+2)=152\) and \(2(36+1+1+1)=78\), respectively.
Both reduce to \(4\) modulo \(37\). Interpolating from
\((0,4),(1,4),(3,29)\) gives \(g_2(T)=4+2T-2T^2\) over
\(\F_{37}\). The prover sends its coefficients, and the verifier then
chooses the next challenge \(r_2\).

Suppose we have a deterministic routine that takes the scopes and tables above and returns \(W\). Write \(A(s,B)\) for
its worst-case running time in bit operations on arity-at-most-three
instances with size parameter \(s\) and weight length at most \(B\). For sparse formulae, \(s = O(n)\), but the reduction works for all \(s\).

\begin{theorem}[From Weighted Counting to a fast honest prover]\label{thm:transfer}
Given a deterministic algorithm for Weighted Counting with bit complexity
\(A(s,B)\) as above, the honest prover for the classical sum-check protocol
for \SAT\ can compute the initial count and all its round
messages in
\[
 O\Bigl((n+m+1)\max_{0\le s\le n+m} A(s,B_p)\Bigr)+\poly(n,m,\log p)
\]
bit operations, where \(B_p\) is the weight bound from
Lemma~\ref{lem:reduction}. The verifier, communication, completeness, and
soundness are those of the classical protocol.
\end{theorem}
\begin{proof}
The prover first computes the claimed count with one call on the instance
with \(N=n\), \(M=m\), and each clause table equal to \(1\) on satisfying settings
and \(0\) otherwise; its value is the exact number of satisfying assignments.
By Lemma~\ref{lem:rounds}, all round messages take at most \(n+3m\) further
calls and \(\poly(n,m,\log p)\) additional bit operations. Every instance has
\(s\le n+m\) and weights of at most \(B_p\) bits, so each call costs at most
\(\max_{0\le s\le n+m}A(s,B_p)\), which gives the stated bound. The prover
sends exactly the messages of the standard honest prover, so the verifier,
communication, completeness, and soundness are unchanged.
\end{proof}

\begin{corollary}\label{cor:subexp}
If \(A(s,B)=2^{o(s)}\poly(s,B)\), \(m=O(n)\), and \(\log p=\poly(n)\), then
the honest prover of Theorem~\ref{thm:transfer} runs in
\(2^{o(n)}\poly(n)\) bit operations.
\end{corollary}
\begin{proof}
Every instance has \(s\le n+m=O(n)\), and \(B_p=\poly(n)\). Substituting these
into the bound of Theorem~\ref{thm:transfer} gives \(2^{o(n)}\poly(n)\).
\end{proof}

\section{Algorithms for Weighted Counting}\label{sec:instantiation}
We now give two algorithms for the Weighted Counting problem defined in Section~\ref{sec:weighted}, using rational interpolation and integer-weighted networks. These are applications of the counting machinery in~\cite{openaieth} and~\cite{claudeeth}.

\subsection{Rational Interpolation}
The first algorithm for the Weighted Counting problem uses the rational
interpolation routine of~\cite{openaieth}.
\begin{lemma}[Weighted Counting via rational interpolation]\label{lem:openai}
For fixed \(k\), the value \(W\) of a Weighted Counting instance with size
parameter \(s\ge 3\) and weights of at most \(B\) bits can be computed
exactly in
\[
 \poly(s,B)\,2^{O_k(s(\log\log s)^2/\sqrt{\log s})}
\]
bit operations and \(\poly(s,B)\) space.
\end{lemma}
\begin{proof}

Each table, \(w_j\), records the values of the \(j\)-th constraint at all Boolean settings of its scope. We replace \(w_j\) by its multilinear extension
\[
 \Phi_j(X_{I_j})=\sum_{u\in\{0,1\}^{I_j}}w_j(u)
 \prod_{v\in I_j:\,u_v=1}X_v\prod_{v\in I_j:\,u_v=0}(1-X_v).
\]
This is a polynomial with integer coefficients, where each coefficient has \(O_k(B)\) bits. 

\bigskip

\emph{Bounded occurrence.} Applying the machinery in~\cite{openaieth} requires that each variable occur in at most \(b\) scopes. In the product in~\eqref{eq:partition}, this need not be the case. We address this by replacing each occurrence of a variable by a fresh
copy. We need to ensure that, across scopes, the values taken by the `same' variable agree, so we join consecutive copies by \(E(y,y')=1-y-y'+2yy'\), the equality
indicator on Boolean inputs. Each variable that occurs in no scope can be deleted if we multiply the final result
by \(2\) for each deleted variable~\eqref{eq:partition}. The new instance has \(N'=\sum_j |I_j|\le kM\) variables and a set \(F\) of
at most \((k+1)M\) constraints; therefore, it has \(O_k(s)\)
constraints. As required by Lemma~4.7 of~\cite{openaieth}, the scope sizes are at most \(k'=\max\{k,2\}\), because each original constraint keeps its scope size (at most \(k\)) when its variables are replaced by copies, and each equality constraint has scope size~\(2\). Secondly, each variable lies in at most \(b=3\) scopes, namely its own original constraint and at most two equality constraints. Since \(\Lambda\) is the sum, over all variables, of the number of scopes containing each variable, and every variable lies in at least one scope and at most three, we get \(N'\le \Lambda\le 3N'\).

\bigskip

\emph{Counting.} Our polynomial has integer coefficients, so we apply Proposition~3.4 of~\cite{openaieth} with \(\mathcal A=\Q\).
It requires the degree
hypothesis~(3.2). In our case, this holds with \(d=\Lambda\le 3N'\) for the following reason: each constraint is multilinear
in its inputs, so its degree is at most the size of its
scope, and the product has total degree at most \(\sum_{h\in F}|I_h|=\Lambda\). Lemma~4.8 of~\cite{openaieth}, with \(\beta=1\), chooses parameters that,
for all sufficiently large \(N'\), satisfy the numerical hypotheses of its
Lemma~4.7. Since our instance satisfies the scope and occurrence bounds checked above,
Lemma~4.7 then builds the design of Definition~3.1 in polynomial time. With
this design, Proposition~3.4 computes \(W\), and the bounds (3.16) and
(4.18) of~\cite{openaieth} show that this takes
\(2^{O(N'(\log\log N')^2/\sqrt{\log N'})}\) operations up to polynomial
factors, provided each operation has polynomial bit cost, which we justify below.

\bigskip

\emph{Bit cost and space.} Each arithmetic operation of the routine costs \(\poly(s,B)\) bit operations, because every number it handles has \(\poly(s,B)\) bits. In Section~3.4 of~\cite{openaieth}, the routine evaluates each constraint only on inputs that are signed sums of entries of the replacement rows \(\vec P_{h,v}\), evaluated at some \(t\in[Q]\) and at \(t+z\) for a formal variable \(z\). By Lemma~4.6 of~\cite{openaieth}, these entries, and hence the inputs, are polynomials in \(z\) whose coefficients are fractions over a common denominator \(d_{\mathrm{den}}\), with \(d_{\mathrm{den}}\) and all numerators of \(\poly(s)\) bits. A constraint is a polynomial of degree at most \(k'\) with integer coefficients of \(O_k(B)\) bits, so its value on such inputs has a denominator dividing \(d_{\mathrm{den}}^{\,k'}\). Multiplying the values of all \(|F|\) constraints gives a denominator dividing \(d_{\mathrm{den}}^{\,k'|F|}\) and numerators of \(\poly(s,B)\) bits. The routine adds at most \(2^{O(s)}\) such products. Since they share this denominator, the sum has numerators with \(O(s)\) extra bits. Appendix~B.4 of~\cite{openaieth} bounds the
bit cost of the final Hermite interpolation by a polynomial. At any time the routine stores polynomially many numbers (Proposition~3.4 of~\cite{openaieth}), which gives \(\poly(s,B)\) space.
\end{proof}

\subsection{Integer-Weighted Networks}
The independent algorithm of~\cite{claudeeth} is formulated directly in terms of integer-valued local tables and equality tests. We use its network definition (Definition~2), grouping lemma (Lemma~8), and the clique recursion of Appendix~A, Theorem~A(b), with the pieces of Appendix~B, Theorem~B. This route gives a sharper exponent for sparse instances than Lemma~\ref{lem:openai}.

\begin{lemma}[Weighted Counting via integer-weighted networks]\label{lem:anthropic}
For fixed \(k\), the value \(W\) of a Weighted Counting instance with size
parameter \(s\ge 2\) and weights of at most \(B\le\poly(s)\) bits can be
computed exactly in
\[
 \poly(s,B)\,2^{O_k(s/\sqrt{\log s})}
\]
bit operations.
\end{lemma}
\begin{proof}
We use Lemma~8 and Theorem~A(b) of~\cite{claudeeth}, applied to the network of its Section~6 with integer tables. Construct a network with one occurrence bit for each incidence of a variable in a constraint; enforce consistency of successive occurrences of each variable by equality tests. A constraint contributes its table weight on its occurrence bits. Each consistent assignment to the occurrence bits corresponds bijectively to an assignment of all occurring variables and contributes precisely its product of weights. Account separately for unused variables and constraints with empty scope.

Apply the clustering of Section~6 and the proof of Theorem~A(c) of~\cite{claudeeth}, replacing the Boolean satisfaction table of each clause by the given integer table. Group at most \(\lceil M/\gamma\rceil\) constraints into each of \(\gamma\) clusters; the setting of a cluster is the list of its occurrence bits, restricted to agree on any equality tests internal to the cluster. Its weight is the product of its local tables. Each occurrence participates in at most two consistency tests. Consequently the setting length and incident shared-bit count are both \(O_k(\lceil M/\gamma\rceil)\), and the total intercluster shared-bit count \(U\) is \(O_k(M)\). The cluster table has entry bit length \(B_c=O(B\lceil M/\gamma\rceil+1)\); evaluating it and the equality-test factorizations takes polynomial time. These parameters satisfy the network definition with polynomial bit-length weights. Crucially, the grouping lemma and Theorem~A(b) allow integer tables, not just zero--one weights.

For clarity about the exponent, use the level parameter \(\kappa=K_\ell\) in the proof of Theorem~A(c), choosing the largest level with \(\kappa\le M^{1/9}\), and take \(\gamma=3\kappa^8\). (Instances below its fixed threshold can be enumerated.) The grouping costs and new clique dimension are independent of the local weights. With the pieces of Theorem~B, \(e=e_\ell=O(1/\sqrt{\log M})\). The bound in Equation~(45) of~\cite{claudeeth} therefore changes only in the logarithmic bit-length term: \(\log(kM)\) inside the \(\kappa^2\) factor becomes \(O(\log(s+B))\), since the grouped tables have integer bit lengths polynomial in \(s,B\). The resulting logarithm of the running time is bounded by
\[
 O_k\bigl(eM+M/\kappa+\kappa^8\log\kappa+\kappa^2\log(s+B)+\log(s+B)\bigr).
\]
For polynomially bounded \(B\), the proof's inequalities \(M^{1/90}\le \kappa\le M^{1/9}\) give \(O_k(M/\sqrt{\log M})\) for the nonpolynomial part (or, uniformly, the weaker \(O_k(s/\sqrt{\log s})\) bound by padding with neutral constraints). Table-query costs and exact integer word operations contribute at most polynomial overhead by the same bit-length accounting in the recursion. This proves the claimed bound, with the unused-variable multiplier computed directly.
\end{proof}

\section{Subexponential Prover for Sparse \SAT}
\begin{theorem}[Subexponential prover for sparse \SAT]\label{thm:main}
Let \(\phi\) be a 3-CNF formula over \(n\) variables with \(m\le cn\) clauses, for a constant \(c\). For any target soundness \(\varepsilon\in(0,1)\) and any prime \(p>\max\{2^n,3m/\varepsilon\}\) with polynomial bit length, the classical sum-check protocol for the exact number of satisfying assignments has perfect completeness, soundness error at most \(3m/p<\varepsilon\), \(n\) rounds, verifier time \(O(n+m)\) field operations, and communication \(O(n+m)\) field elements. Assuming the algorithmic results in Lemma~\ref{lem:anthropic}, its honest prover runs in
\[
 \poly(n,\log p)\,2^{O_c(n/\sqrt{\log n})}
\]
bit operations. The alternative instantiation in Lemma~\ref{lem:openai} gives \(\poly(n,\log p)\,2^{O_c(n(\log\log n)^2/\sqrt{\log n})}\) bit operations and polynomial space.
\end{theorem}

\begin{proof}
Every evaluation in Lemma~\ref{lem:reduction} has \(s\le n+m=O_c(n)\) and weights of polynomial bit length in \(n\). If needed, pad a late-round instance to size at least \(n\) with fresh Boolean variables carrying the unary indicator \(1-x\), whose sum is one. This preserves the weighted sum and makes the bit length polynomial in the padded instance size. Apply Lemma~\ref{lem:anthropic} or Lemma~\ref{lem:openai} in Theorem~\ref{thm:transfer}. The unchanged sum-check analysis in Section~\ref{sec:prelim} establishes all verifier-side guarantees.
\end{proof}

\begin{center}
\begin{tabular}{@{}llll@{}}
\toprule
Honest prover & Time (sparse case) & Verifier & Rounds\\
\midrule
Standard & \(\poly(n)\,2^n\) & \(O(n)\) & \(n\)\\
Rational interpolation & \(\poly(n)\,2^{O(n(\log\log n)^2/\sqrt{\log n})}\) & \(O(n)\) & \(n\)\\
Integer-weighted networks & \(\poly(n)\,2^{O(n/\sqrt{\log n})}\) & \(O(n)\) & \(n\)\\
\bottomrule
\end{tabular}
\end{center}
Here \(\log p=\poly(n)\); the prover bounds count bit operations and the verifier column counts field operations. For example, with a prime of \(O(n)\) bits, communication is \(O(n^2)\) bits and the verifier's bit complexity is polynomial. This distinction matters when interpreting the linear bounds.

The sharper prover bound lies on the same asymptotic exponent scale as the stated sparse Boolean counting algorithm. This is a doubly efficient proof in the sense of obtaining a prover time polynomial in a specified subexponential solver upper bound, rather than a proof of optimality or a pointwise polynomial-factor overhead over the fastest possible solver. In particular, hidden constants in the exponents may differ. Neither counting preprint presently certifies a useful practical speedup at realistic instance sizes.

\section{Related Work}



A recent alternative proof of the ETH refutation by Mazor and Pass~\cite{mazor} uses geographic information retrieval and may offer a further approach. A systematic comparison with that construction and other ongoing post-ETH work is left for revision. Our preliminary search did not identify an earlier explicit statement of Theorem~\ref{thm:main}; this is not a priority claim. Since the Weighted Counting problem is already present in the ETH-refutation preprints, the present result should be understood as an application and specialization of that machinery.


\section{Future Work}\label{sec:open}
For a dense 3-CNF, Theorem~\ref{thm:transfer} only gives a bound in terms of \(s=n+m\), not necessarily subexponential in \(n\). The Boolean counting algorithms overcome this issue by replacing \(\phi\) with formulae
that have the same total count, through sparsification and algebraic reductions. The
sum-check prover cannot: its messages are values of the fixed polynomial
\(g\) built from \(\phi\), and changing the formula changes \(g\) even when
the truth table is unchanged. For example, \(\phi\wedge\phi\) has the same
truth table as \(\phi\), but its arithmetization is \(g^2\neq g\).

An important question is whether there is a direct \(2^{o(n)}\poly(n,m,\log p)\)-time algorithm for precisely the restricted weighted sums in Lemma~\ref{lem:reduction} for arbitrary-density 3-CNF. The weights have more structure than arbitrary tables, since they arise as partially evaluated extensions of clauses; exploiting this structure could be useful. 

Further directions include reducing the prover's overhead over a single run
of the counting algorithm (currently \(O(n+m)\) calls), fewer rounds,
small-space implementations, and extensions to other local constraint
systems. Theorem~\ref{thm:transfer} already extends verbatim to bounded-width CNF and, more generally, products of bounded-arity polynomial factors with polynomial degree and efficient local evaluation. Whether stronger results hold for arbitrary Boolean formulae, rather than bounded-width CNF, remains separate from the sparse result proved here.

\section{Acknowledgments}\label{sec:ack}
We used OpenAI’s ChatGPT and Anthropic’s Claude to explore proof ideas and refine the exposition. The authors take full responsibility for the paper’s content.

\bibliographystyle{plain}
\bibliography{references}
\end{document}